\documentclass[preprint2]{aastex63}

\submitjournal{PSJ}

\shortauthors{Denman et al.}

\graphicspath{{./}{figures/}}

\begin{document}
\title{The Influence of Temperature and Photobleaching on Irradiated Sodium Chloride at Europa-like Conditions}
\correspondingauthor{William T.P. Denman}
\email{wdenman@caltech.edu}
\author[0000-0003-4752-0073]{William T.P. Denman}
\affil{Division of Geological and Planetary Sciences, California Institute of Technology, Pasadena, CA 91125, USA}\author[0000-0002-0767-8901]{Samantha K. Trumbo}
\affiliation{Division of Geological and Planetary Sciences, California Institute of Technology, Pasadena, CA 91125, USA}
\affiliation{Cornell Center for Astrophysics and Planetary Science, Cornell University, Ithaca, NY 14853, USA}
\author[0000-0002-8255-0545]{Michael E. Brown}
\affil{Division of Geological and Planetary Sciences, California Institute of Technology, Pasadena, CA 91125, USA}

\begin{abstract}
Europa's leading hemisphere chaos regions
have a spectral feature at 450 nm
that has been attributed to absorption by
crystal defects
in irradiated sodium chloride, known as F-centers.
Some discrepancies
exist between the laboratory data of irradiated
sodium chloride and the observations, including
a $\sim$10 nm shift in central wavelength of the F-center band 
and the lack of the prominent
720 nm absorption on Europa from M-centers, which result from the coalescence 
of pairs of F-centers.
Here, we perform irradiation experiments on sodium
chloride in an attempt to understand
these discrepancies. We show that careful
control of the temperature of the sample at
a temperature of 120 K yields
F-centers with an absorption wavelength 
comparable to that of Europa. In addition,
we measure the effect of photobleaching -- the destruction
of F-centers by photons -- and show that at
the energetic particle and photon flux on
Europa, an equilibrium will be reached 
where only a modest F-center absorption
develops. The density of F-centers never reaches
high enough values for the creation of
secondary M-centers. Our experiments predict
that F-centers grow during the night on Europa
in the absence of photobleaching and then partially decay 
during the daytime. We show observations from
the Hubble Space Telescope consistent with
this prediction. All observations of the
450 nm F-center on Europa are now consistent with
laboratory measurements of sodium chloride,
confirming the presence of this salt on Europa.
\end{abstract}


\section{Introduction} \label{sec:intro}

Jupiter's moon Europa has a young, icy surface which undergoes high energy radiation bombardment and covers a salty liquid water ocean \citep{kivelson_galileo_2000, paranicas_electron_2001, paranicas_ion_2002}. The salts in the ocean presumably derive from the interaction between
the liquid water and the silicate seafloor \citep{kargel_europas_2000}, and
the composition of the salts offer insight into the manner of this interaction. While no direct access to the ocean is currently
possible, Europa's geologically young chaos regions have likely been resurfaced by
material ultimately derived from the interior ocean
and thus could provide information on its composition \citep{carr_evidence_1998, glein_sodium_2010,schmidt_active_2011,soderlund_ocean-driven_2014}.
On the trailing side of Europa these chaos terrains are 
bombarded by sulfur plasma, confusing the interpretation
of their initial composition; the leading side chaos terrains
are thus the most likely to retain signatures of the original
composition. 

\citet{trumbo_sodium_2019} recently reported {\it Hubble Space
Telescope} (HST) observations that showed the existence of 
an absorption feature at 450 nm in Europa's leading-side
chaos terrains and proposed that this feature is due to sodium chloride 
salts. While salts are generally indistinct in the visible
wavelength range, irradiation of alkali halides, in particular,
produce multiple defects and dislocations within the crystal
structure, giving rise to distinct ``color center'' absorptions.
For NaCl, the most prominent color center is the F-center, 
occurring at approximately 450 nm \citep{mador_production_1954}.

F-centers occur when radiation ejects a chlorine atom
from its location in the crystal lattice and an electron
fills the hole. More complicated color centers also form, 
including  M-center absorption at 720 nm, which appear
when the density of F-centers is sufficiently high that pairs
begin to coalesce. This F-center coalescence can eventually 
lead to macroscopic sodium colloids which broadly absorb in the
580 nm region. F-centers can be destroyed by absorption
of photons within the 400-500 nm band of the absorption 
feature in a process called photobleaching, which
excites the electron into the conduction band \citep{herman_optical_1955}.

Experiments performed in order to understand
the possibility of NaCl color centers on Europa \citep{hand_europas_2015, poston_spectral_2017, hibbitts_color_2019} suggested that
irradiation at conditions similar to
those on Europa would form
both F-centers and M-centers. The HST data, however,
did not completely match the laboratory data. In the laboratory,
the central wavelength of the F-center was measured to be 460 nm,
rather than the $\sim$450 nm seen on Europa, and the HST data showed
no hint of the predicted absorption near the 720 nm M-center.
\citet{trumbo_sodium_2019} noted that when the irradiation of the NaCl
stopped in the \citet{poston_spectral_2017} experiments, the F-center
wavelength quickly shifted blueward as the absorption rapidly
decayed. They hypothesized that irradiation at the lower 
flux level
seen on Europa could lead to the shorter wavelength feature. 
They further hypothesized that the lack of M-centers on Europa
is the result of solar photobleaching which would destroy F-centers
at nearly the same rate as their creation through irradiation. This destruction would lead to an equilibrium with a low number of F-centers that would not have a high enough density for M-centers to form.

Here, we perform experiments to test these hypotheses. Our goal
is to determine if the differences in appearance between the HST
spectra and the laboratory data can be reconciled with the
interpretation that the $\sim$450 nm absorption feature 
seen on Europa is indeed due
to irradiated NaCl.

\section{Method} \label{sec:method}
Our experiments were designed to measure the spectral changes in sodium chloride from electron and photon bombardment at temperatures similar to Europa. For each experiment, fresh hydrated sodium chloride powder, USP (CAS-7647-14-5) was pressed into a layer of indium foil inside of a copper sample cup. Excess salt was brushed off to ensure good thermal coupling 
between the remaining grains and the sample cup.  
The sample cup was then 
loaded onto a cold finger
inside of a Kimball Physics ultrahigh vacuum chamber. The chamber was pumped down to $~10^{-8}$ torr using an Agilent ID3 backing pump and TwisTorr 84 molecular turbo pump. The cold finger was cooled
with a Janis ST-400 liquid nitrogen cryostat. Once the desired temperature was reached, the grains were irradiated with 10 keV electrons from a Kimball Physics EGG301 electron gun. The current is measured with a Kimball Physics Faraday cup mounted
on a linear actuator that can be inserted into the electron beam
path.  Electron currents used in the
different experiments and their relationship to the approximate energy
flux on Europa are shown in Table \ref{table1}. We obtained spectra of the samples from 300-1000 nm by illuminating the sample via an external
lens fed by a fiber optic connected
to both a stabilized deuterium and a stabilized tungsten-halogen lamp from Thorlabs. We illuminated the sample at 45 degrees from the surface normal and collected the diffuse reflection 90 degrees from the specular direction and 45 degrees from the surface normal. Since NaCl is nearly featureless through the spectral range, unirradiated salt at the desired temperature was used as a spectral reference for the our samples. 

\begin{deluxetable*}{cccccc}
\label{table1}
\tablenum{1}
\tablecaption{Experiments\label{table:experimental}}
\tablewidth{0pt}
\tablenum{1}
\tablehead{
\nocolhead{} & \colhead{current density} & \colhead{electron flux} & \colhead{450 nm irradiance} & \colhead{dose rate}  & \colhead{temperature}\\
\nocolhead{} & \colhead{(nA/cm$^2$)} & \colhead{(electrons/cm$^2$/s)} & \colhead{(mW/m$^2$/nm)}  & \colhead{(eV/16amu/yr)}  & \colhead{(K)}}

    \startdata
        Low  & $12$ & $8 \times 10^{10}$ & $6.0$ & $2.5 \times 10^{3}$   & $120$\\
        Low & $12$ & $8 \times 10^{10}$ & $-$ &  $2.5 \times 10^{3}$ & $120$\\
        $-$ & $48$ & $3 \times 10^{11}$ & $6.0$ & $1 \times 10^{4}$ & $120$\\
        Medium & $505$ & $3 \times 10^{12}$ & $6.0$ & $1 \times 10^{5}$  & $120$ \\
        Medium & $505$ & $3 \times 10^{12}$ & $-$ & $1 \times 10^{5}$  & $120$\\
        Medium & $505$ & $3 \times 10^{12}$ & $6.0$ & $1 \times 10^{5} $& $120$(no indium foil)\\
        Medium & $505$ & $3 \times 10^{12}$ & $-$ & $1 \times 10^{5}$  & $200$\\
        High & $3410$ & $2 \times 10^{13}$ & $6.0$ & $7 \times 10^{5}$  & $120$\\
        High & $3410$ & $2 \times 10^{13}$ & $-$ & $7 \times 10^{5}$ & $120$\\
        Europa & n/a & n/a & $78$ & $10$ & $120$ \\
    \enddata
\tablecomments{The energy deposited is for the top 1.2 $\mu$m micron of the sample. For the experiment without indium foil the temperature of the salt is unknown. The irradiance
of our spectral lamp is estimated at 450 nm. For Europa,
the photon flux and temperature are given for equatorial noon.}
\end{deluxetable*}
 \begin{figure}[h!]
    \figurenum{1}
    \plotone{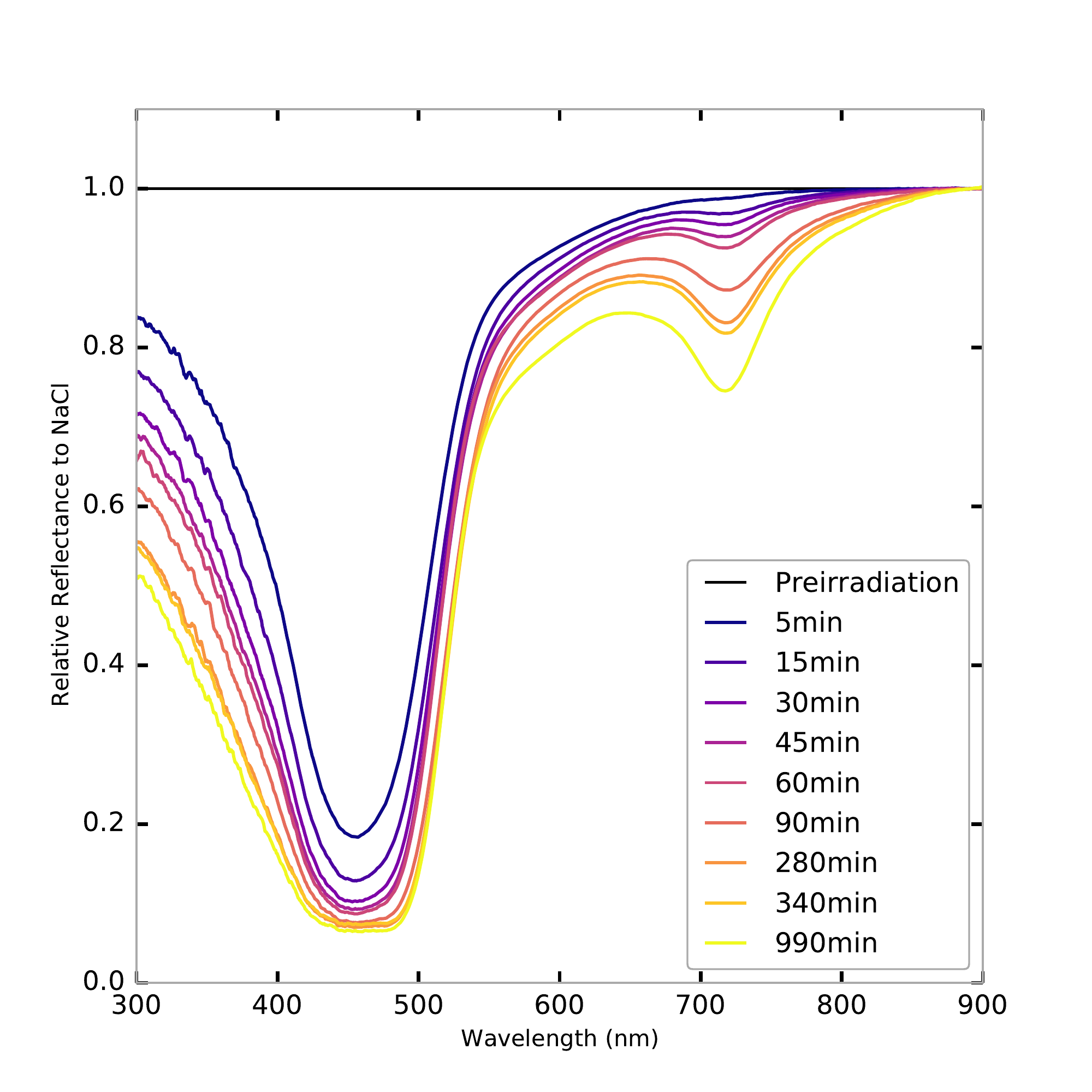}
    \caption{Spectra of irradiated sodium chloride at 120K with our high irradiation dosage
    and no photobleaching. M-center growth is observed after F-center saturation.}
    \label{fig:0512high}
\end{figure}
For our series of experiments the electron flux and illumination conditions were varied. In our initial experiments, we simply
tested the system to ensure that we could create and measure
F- and M-centers. Our initial electron flux, as
measured in the Faraday cup, was $2\times 10^{13}$ electrons cm$^{-2}$ s$^{-1}$. Since the penetration depth of 10 KeV electrons is around 1.2$\mu$m \citep{hand_europas_2015} such a flux gives 
the equivalent of $7\times 10^5$ eV (16 amu)$^{-1}$ deposited each year
into the NaCl. The energy deposited on the apex of Europa's leading hemisphere is estimated to be 10 eV (16 amu yr)$^{-1}$ \citep{nordheim_preservation_2018}. Our initial experiments are at a dosage rate approximately 70000 times that of Europa. The true dosage rate on Europa is both poorly determined and variable, so this ratio is only an approximation. Additional experiments were run at fluxes 7 times lower and nearly 300 times lower than our initial experiment. These three settings are referred to as high, medium and low flux. It should be noted that even the low flux is $\sim250$ higher than that of Europa.
Figure \ref{fig:0512high} shows the results of our first experiment at high
flux. F-centers are quickly created and become saturated within
approximately 30 minutes. As the density of F-centers increases,
coalescence of these defects begins to form M-centers, which
begin to absorb at 720 nm by 30 minutes. By the end of the
experiment both F- and M-center absorptions are prominent.

In some experiments, we test the effects of photobleaching
by leaving our spectral lamp illuminated during the
irradiation. While we have no means of measuring the
absolute flux density of our spectral lamp inside of the chamber
at 450 nm, we estimate this 
parameter by using the spectral curve and total output
provided by the manufacturer as well as the illumination
spot size on the sample. We find the irradiance of the lamp at 450 nm to be 6 mW m$^{-2}$ nm$^{-1}$. This irradiance is approximately $8\%$ of the flux density at Europa from solar illumination.

\section{Experiments} \label{sec:Experiments}
 The visible spectrum of the leading side chaos regions on Europa differs from laboratory spectra of irradiated NaCl in two important ways. First, the Europa feature appears at 
 $\sim$450 nm, rather than the 460 nm wavelength found for NaCl F-centers in the laboratory \citep{poston_spectral_2017,hand_europas_2015}.  Second, the laboratory spectra all show the formation of an M-center at $\sim 720$ nm, which the HST data strongly rule out. We perform a set of experiments to explore these phenomena.  
 \begin{figure}[h!]
    \figurenum{2}
    \plotone{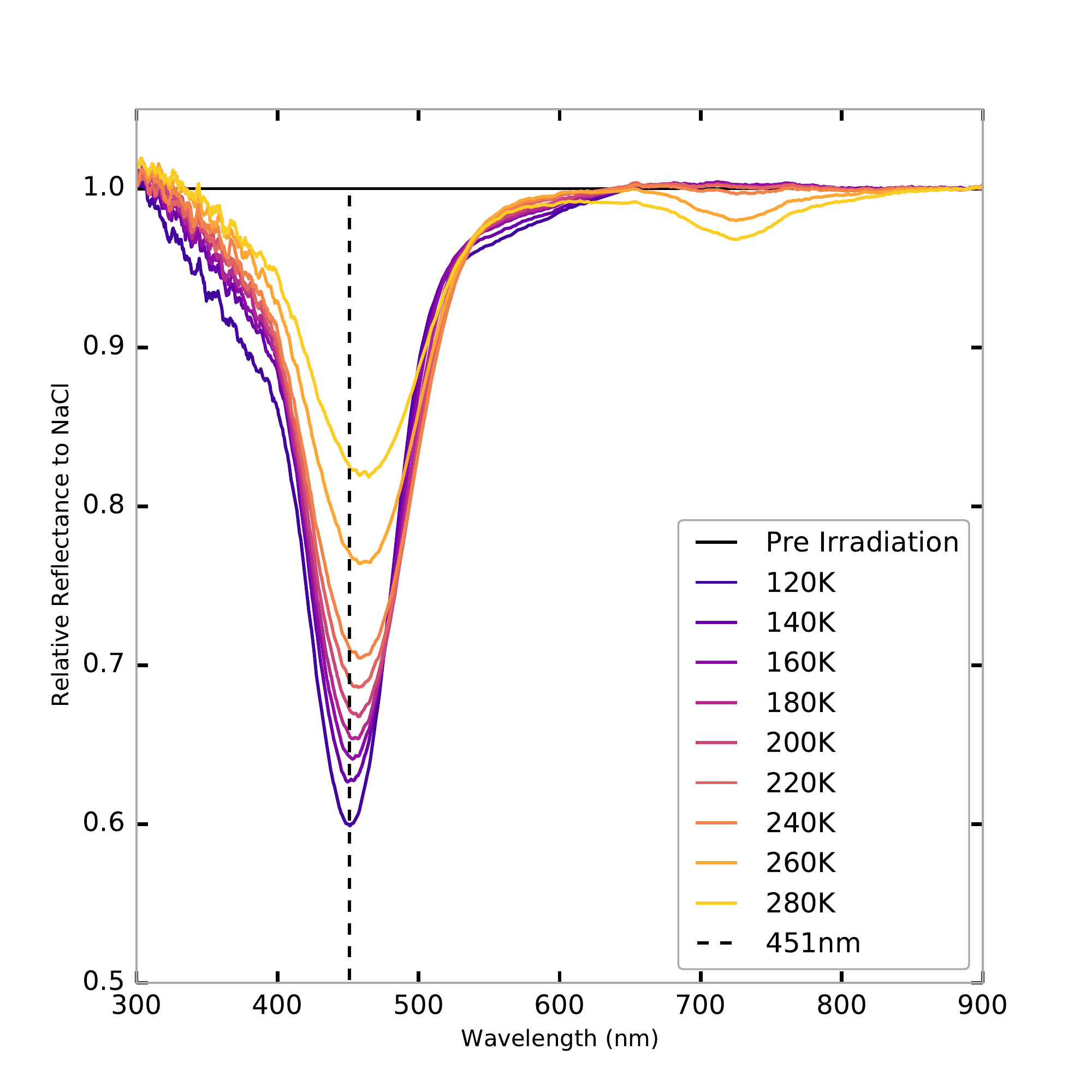}
    \caption{Irradiation at 120K with systematic warming of the sample. This shows the redward shift of the F-center with temperature. The central wavelengths from 120K to 280K are 451.2 nm, 453.1 nm, 454.1 nm, 456.1 nm, 457.6 nm, 459.0 nm, 460.5 nm, 461.9 nm, 463.4 nm. The minimum values were found by doing a parabolic fit to the F-center range (400-500 nm) and then finding the wavelength minimum of the parabola.}
    \label{fig:120kwarmup}
\end{figure}
\subsection{Wavelength Shift}
 \citet{trumbo_sodium_2019} noted that, in experiments by \citet{poston_spectral_2017}, the F-center spectral feature quickly decayed and shifted blueward when irradiation ceased. They hypothesised that the higher irradiation flux for the laboratory spectra caused the mismatch in wavelength between the Europa and the laboratory spectra. Our first set of experiments were designed to test that hypothesis.
  
 We first attempted to reproduce some of the key results of \citet{poston_spectral_2017}. We irradiated a NaCl sample at a temperature of 120 K at our medium flux level, which is approximately 6 times
 that of the initial \citet{poston_spectral_2017} irradiation, with no photobleaching. This flux is approximately $10^4$ times the flux level at Europa. We irradiated for 2 hours, periodically obtaining spectra, to measure the growth of both the F-center and the M-centers. This 2 hour irradiation is equivalent to 2.3 years on Europa. We observed a central wavelength for the F-center of 451 nm, which is significantly shifted from the \citet{poston_spectral_2017} results and much closer to the spectra from Europa (Figure \ref{fig:120kwarmup}), even though our flux level was 6 times higher.
 At the end of the irradiation sequence, the samples were left in the dark at 120K overnight. In contrast to \citet{poston_spectral_2017}, we observed no change in the spectrum after ceasing irradiation. We then systematically warmed the sample, allowing the sample to stabilize at each new temperature, and obtained a spectrum at each, in order to examine the effects of temperature. As noted by \cite{schwartz_effect_2008}, the wavelength of the F-center systematically shifts redward with increasing temperature. As well as shifting, the depth of the F-center decreases with increasing temperature. By 240 K the peak of
 the F-center absorption is at 460 nm,
 and at 260 K, M-centers
 begin to spontaneously grow, as the higher temperatures allows
 the F-centers to mobilize and begin to coalesce (Figure \ref{fig:120kwarmup}).

\begin{figure}[h!]
\figurenum{3}
\plotone{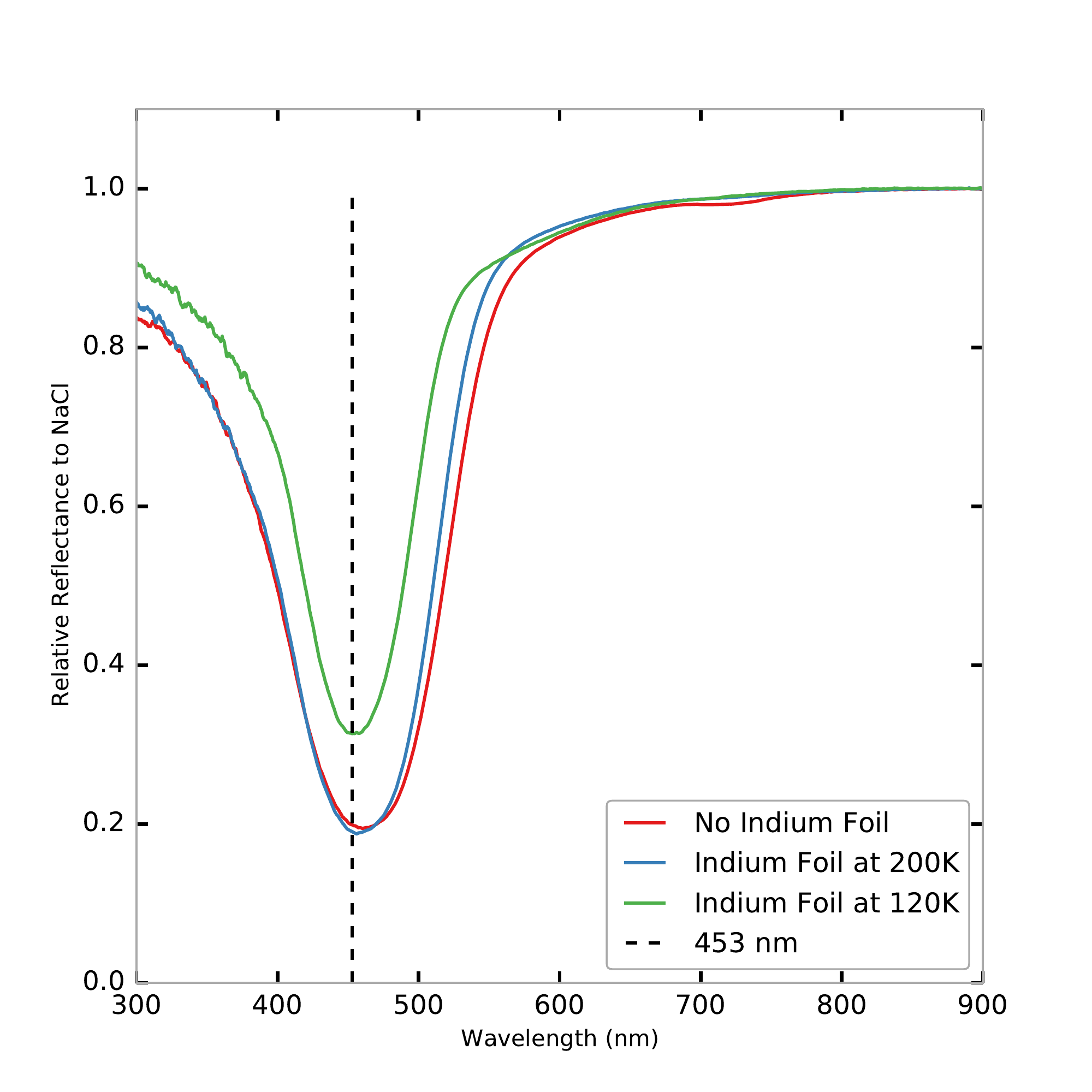}
\caption{Spectra of NaCl irradiated at medium irradiation for 15 minutes. Significant differences are seen in samples with indium foil at 120K, and 200K, and without indium foil.
F-center growth is significantly faster for the 200 K and no-indium foil samples, while the F-center wavelength is also shifted redward. For the no-indium foil sample, M-center
growth can already be seen. We conclude that the sample
sitting on a 120 K cold finger but without indium foil to
thermally couple it is even warmer than 200 K. 
\label{fig:noindium}}
\end{figure}

The difference between these experiments and the results of 
\citet{poston_spectral_2017} are striking. We obtain a significantly shorter
wavelength for the F-center, and we see no spontaneous
spectral change post-irradiation until we purposely raise the
temperature. We hypothesize that these differences are due to a lack
of adequate thermal coupling between the salt
grains and the cold finger in the \citet{poston_spectral_2017} experiments. The angular grains have little contact with each other or with the cold finger and are exposed to thermal emission from the walls of the experimental chamber. These grains could be at a significantly higher temperature than the copper sample cup and the cold finger. By embedding the samples in indium foil, as was done in the present experiments, the thermal coupling is greatly increased. To test this theory, we irradiated salts in the same manner as before, but without pressing the sample into indium foil. We found that the central wavelength shifted to 460 nm, and the growth of both the F-center and M-center occurred on much faster time scales when compared to spectra at 120K
(Figure \ref{fig:noindium}).  As a final test, we irradiated indium-pressed thermally coupled salt grains at 200K. For these spectra the central wavelength was observed at 458 nm. There was no discernible shift in the central wavelength post irradiation but a slight growth in the M-center was observed post irradiation. The peak of the absorption
for these 200K experiments was blueward of both the \citet{poston_spectral_2017} spectra as well as our experiments without pressing the salt into the indium foil (Figure ~\ref{fig:noindium}). Based on these experiments, we hypothesis that the grains in the \citet{poston_spectral_2017} experiments were not well coupled to the cold finger. We believe they were approximately $\sim$ 240K, based on the central wavelength and the growth of the M-center. This higher temperature would naturally lead to the redward shift of the F-center and the post-irradiation spectral changes observed in those experiments but not observed in our thermally coupled 120 K experiments.  We conclude that the shift in the F-center between Europa and laboratory data is not due to the increased irradiation in the laboratory, but rather from the inadvertently elevated sample temperature in the previous laboratory data. For thermally coupled samples at 120K, the central wavelength of 451 nm is within the uncertainty of the positional measurement of the feature on Europa. The position we observed also coincides with the central wavelength previously measured for these temperatures \citep{mador_production_1954,schwartz_effect_2008}.
\subsection{M-Center Formation}
\begin{figure}[h!]
\figurenum{4}
\plotone{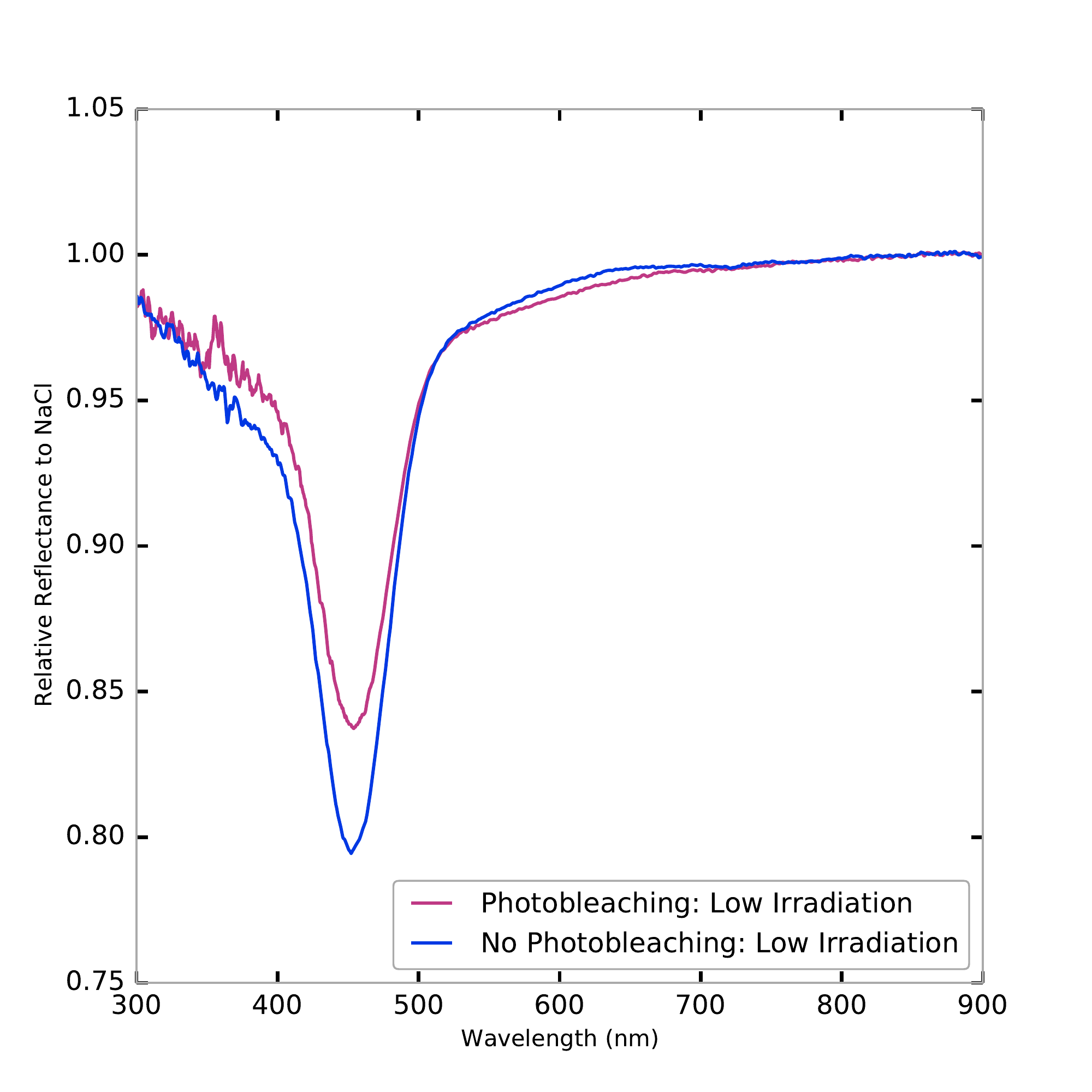}
\caption{Spectrum of NaCl irradiated at low irradiation for 40 minutes with and without photobleaching. The central wavelength is the same for both spectra.
\label{fig:photobleachlow}}
\end{figure}

\begin{figure}[h!]
\figurenum{5}
\plotone{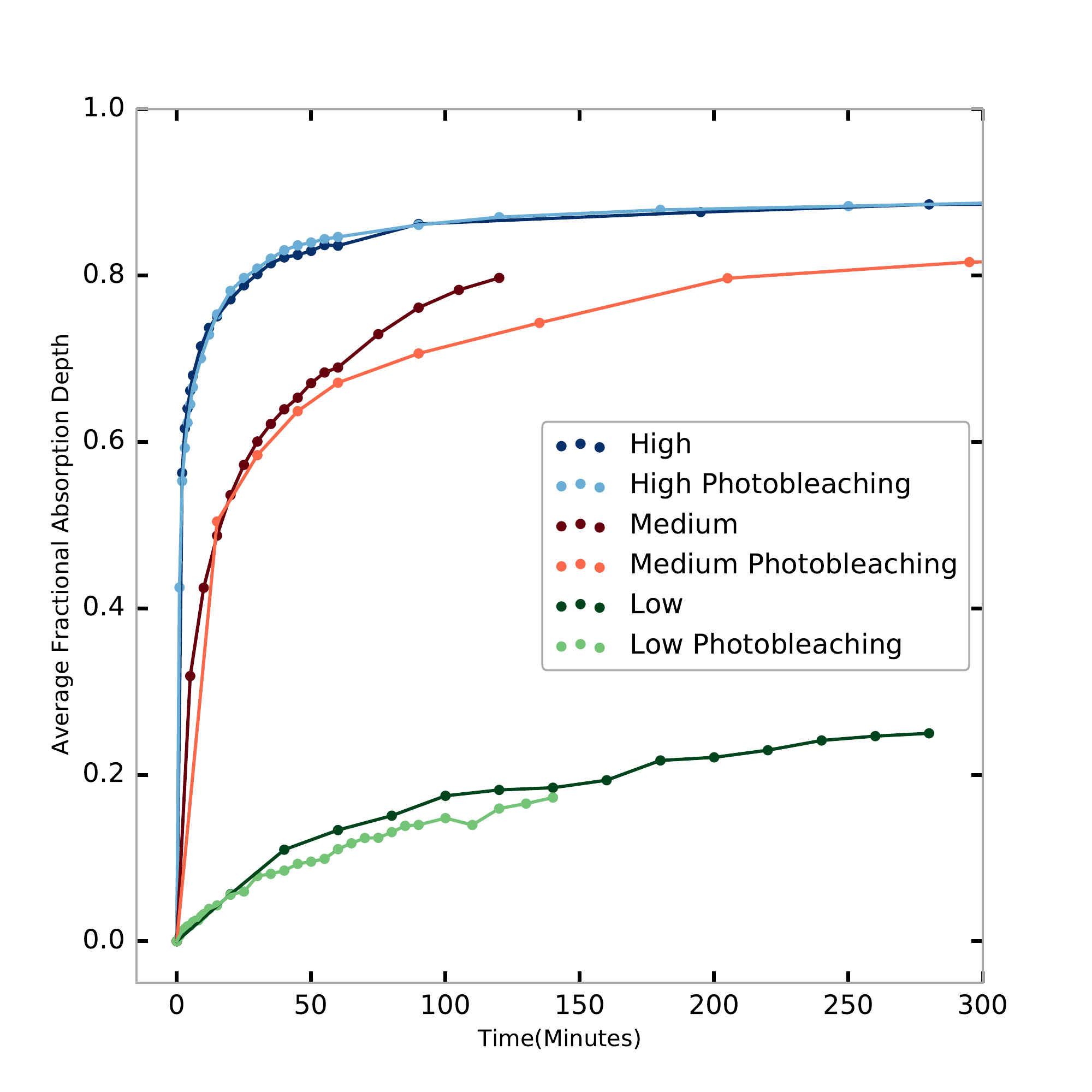}
\caption{The average fractional absorption depth of the F-center from 425-475 nm  vs time of irradiation for
our three irradiation levels, with and without photobleaching. In all cases, photobleaching slowed the growth
of the F-centers, though the lamp is too weak to completely balance even the lowest electron irradiation level.
\label{fig:areavstime}}
\end{figure}

The second critical difference between
the laboratory and HST data
is the strong growth in the laboratory
of the M-center absorption feature
located at 720 nm
compared to its non-detection from HST.
\citet{trumbo_sodium_2019} hypothesized that photobleaching, the destruction of F-center color centers by photons, could suppress the formation of F-centers
sufficiently that M-centers, which require a certain density of
F-centers in order to coalesce, might never form. 
To test the magnitude of this photobleaching effect, we performed two sets of experiments. The first was a series of experiments with and without photobleaching at three different irradiating levels. 
For photobleaching we use the same lamp as is used for spectral reflection. Our lamp provides roughly 8\% of the photon flux density in the 400 to 500 nm wavelength span of the F-center compared to noon at Europa's equator.  At all irradiation levels, photobleaching slowed the growth of the F-centers, as can be seen in the comparison between photobleached
and non-photobleached spectra with otherwise identical conditions in Figure \ref{fig:photobleachlow}. To additionally examine this effect, we measured the mean absorption depth of the F-center
from 425-475 nm as a function of time for each of our
irradiation experiments.  
As seen in Figure \ref{fig:areavstime}, the average absorption depth of the F-center is dependent on both electron flux and whether or not there is photobleaching. The growth rate of the F-centers, without photobleaching, is approximately linear
until saturation occurs and the formation of M-centers begins.
This linear region is poorly resolved in the medium and high
irradiation experiments, but if we take the 5.6\% absorption
growth in the first 20 minutes of the low irradiation
experiment to be typical
of the quasi-linear growth regime, we find a growth
rate of 17\% hr$^{-1}$. Scaling by the irradiation flux
gives consistent results for the medium and high irradiation
levels. We thus assume that for small F-center absorption depths, the
growth of the absorption depth,  $A$, can be described as a simple linear
growth, $dA/dt=aF$. Here, $F$ is the irradiation flux in units
of eV $(16{\rm amu} {\rm yr})^{-1}$, where the flux at Europa is
$\sim$10, and $a$ is a constant, here found to
be $6.8\times 10^{-5}$ in the same units as $F$ and with $t$
in hours. At the flux of Europa, a $\sim$7\% absorption feature
would grow in a little more than one orbital period in
the absence of competing effects.

It is difficult to simultaneously match the 
photon and electron flux at Europa
with our experimental setup. Even our lowest level of irradiation 
is $\sim$250 times that experienced at Europa, while our
photobleaching photon flux density is $\sim$13 times {\it less}. 
Moreover, every time we take a spectrum we are photobleaching 
for the $\sim$1 minute duration of the spectral exposures. Thus
even the unbleached spectra have actually had some amount of
photobleaching.
For these reasons, it is difficult to extrapolate the results of
Fig. \ref{fig:areavstime} to estimate the effect at Europa.

\begin{figure}[h!]
\figurenum{6}
\plotone{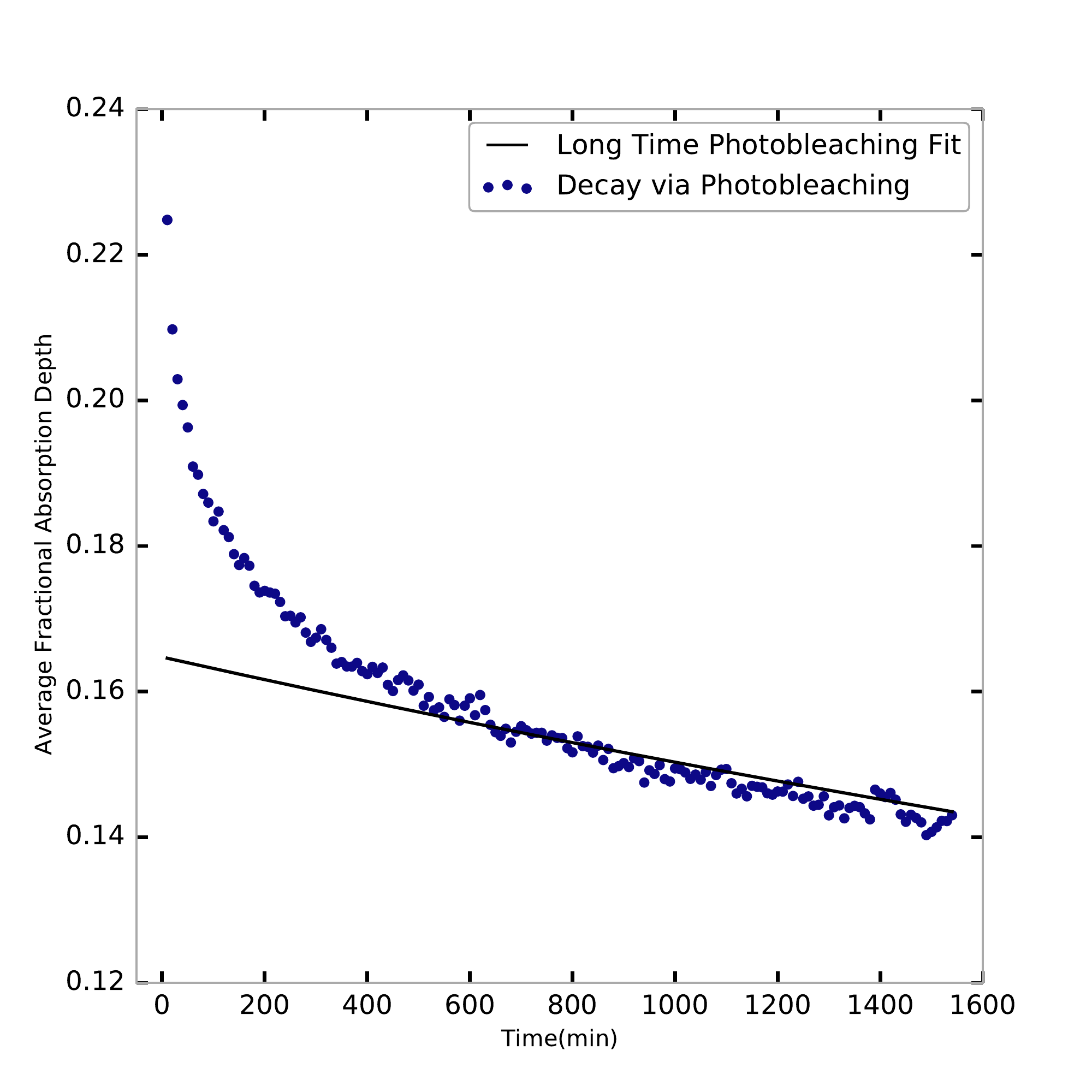}
\caption{The average fractional absorption depth of the F-center from 425 to 475 nm with photobleaching post irradiation.
\label{fig:pbfcenterarea}}
\end{figure}

To attempt to obtain a better estimate of the effect of
photobleaching on Europa, we performed one
additional experiment. We irradiated a sample at four times
our low irradiation flux for 10 minutes, 
leading to an F-center with a fractional absorption depth of about 20\%,
approximately three times that seen on Europa.
We than ceased irradiation and began photobleaching. 
The F-center absorption decayed rapidly for the first few hours
(Fig. \ref{fig:pbfcenterarea}).
By the time it reached values more comparable to those
seen on Europa, the decay slowed, and we fit this
section of the curve
to the theoretically derived and experimentally
verified equation for photobleaching of
lightly irradiated NaCl, 
$dA/dt=-bIA^2$, where $A$ is the fractional absorption depth of the F-center,
$I$ is the intensity of the light, and $b$ is a constant.
\citep{herman_optical_1955}. We find a value of 0.035 for the product $bI$ for our lamp, which scales to 0.46 for the noon time equatorial solar flux at Europa.

\begin{figure}
\plotone{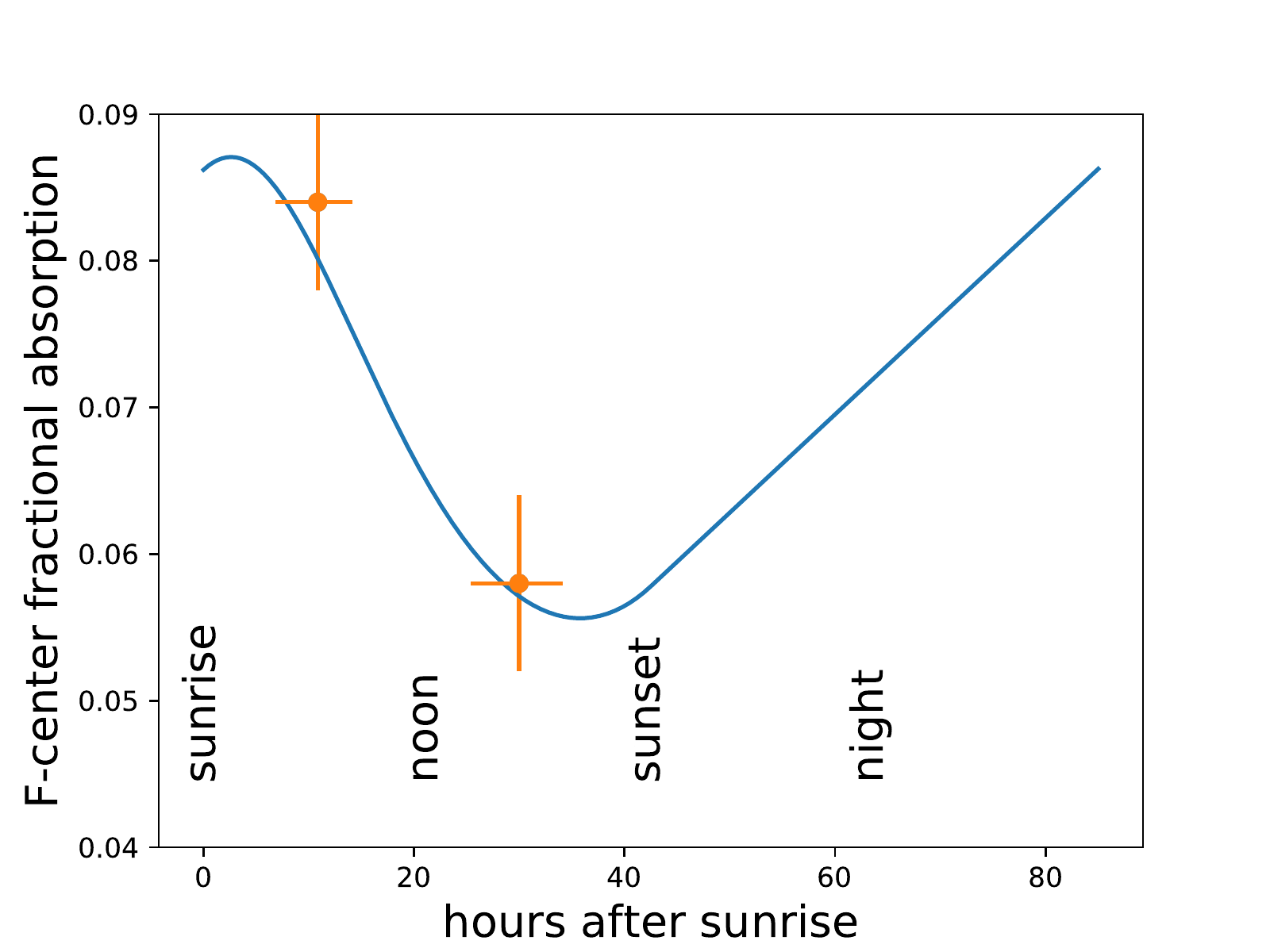}
\figurenum{7}
\caption{A simple numerical model showing the F-center
fractional absorption as a function of Europa's 85 hour
rotation period, including the effects
of continuous irradiation on Europa with solar flux
dependent photobleaching during the daytime. The F-center
grows over the course of the night and begins decaying soon
after sunrise before recovering after sunset. The data points
show the absorption depths of
two HST measurements of identical regions within
Tara Regio taken at different local times. The horizontal
error bars show the geographic extent of the observed region.
The F-center 
decay predicted by the model is observed on Europa.
\label{fig:numpb}}
\end{figure}

We used this photobleaching equation and an assumed linear growth
rate of F-centers with irradiation to estimate an equilibrium
value of F-center absorption depth. We model the F-center growth rate
as $dA/dt=aF-bIA^2$.
In steady-state and assuming continuous photobleaching
at the average equatorial day/night irradiance, 
this equation predicts an absorption
depth of 6.8\% for Europa. This calculated value is remarkably similar to
the depth seen in the HST data.
Crucially, such an equilibrium 
value for the F-center is much lower than that seen to be
required for formation of M-centers.
In reality,
photobleaching is not constant on Europa, but ranges
from zero during the night
to a maximum at noon. Incorporating this changing photobleaching
rate into a simple numerical model which accounts
for nighttime and sinusoidally varying illumination in
the daytime suggests that, in equilibrium, the
F-center absorption depth would grow to 8.7\% overnight, decay to 5.6\% by the end of
the day, and regrow overnight (Figure \ref{fig:numpb}). 

Intriguingly, the HST data on Europa show evidence for this precise photobleaching
effect. While \citet{trumbo_sodium_2019} report only average values for
the F-center strength across the disk, we note that a portion of
Tara Regio was observed both in the morning and in the
afternoon. Reproducing their data reduction, but examining
overlapping spectra separately, we find that the morning absorption depth 
is indeed deeper than that of the afternoon. These absorption depths are well fit by our simple time-dependent mode.
Given the crudeness of
the modeling and the difficulty of extrapolating laboratory 
conditions to those at Europa, we believe that the extremely
close agreement between the precise values of the
model and of the data must be mainly coincidental. Nonetheless,
we find the fact that the relative change
in the F-center absorption depth from HST agrees with the model
to be an encouraging sign that our extrapolations to Europa's
surface conditions and crude modeling are capturing 
the real behaviour at Europa.

\begin{figure}[h!]
\figurenum{8}
\plotone{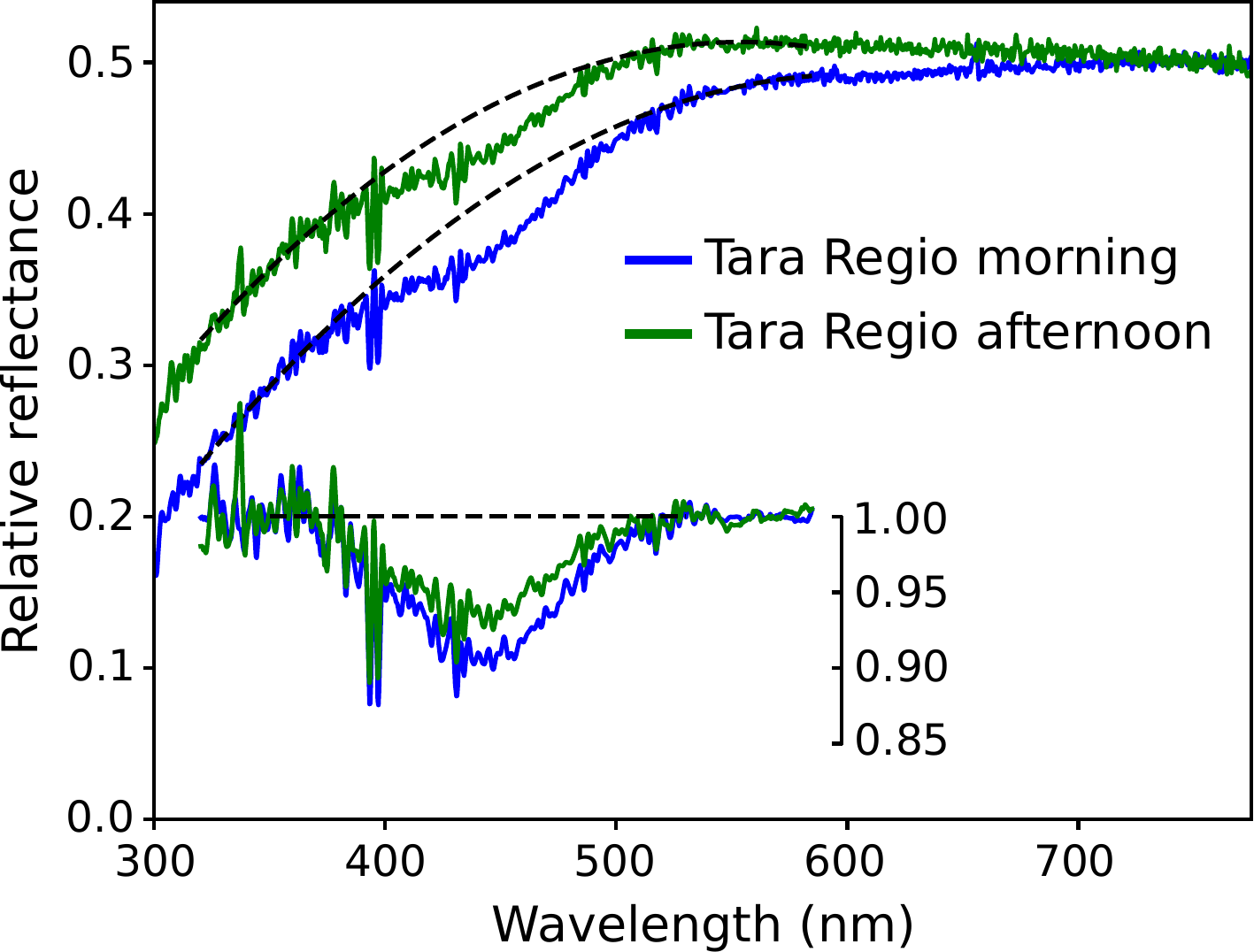}
\caption{HST observations of overlapping regions of
Tara Regio in the morning and in the afternoon. Dashed lines are second-order polynomial continuum fits, which facilitate band-strength comparison. The continuum-removed absorption bands are shown
below. The observed
decay in the F-center absorption is consistent with the
photobleaching decay predicted here.}
\end{figure}

\section{Conclusions}\label{sec:conclusions}
We used new laboratory data with carefully controlled temperature
and photon illumination conditions to examine the discrepancy
between the spectrum of the Tara Regio region
on Europa measured with HST and the spectrum of irradiated
NaCl found in previous laboratory work. 
We found that in our experiments irradiated NaCl grains,
which are thermally coupled to a 120 K cold finger by 
pressing them into indium foil, develop
F-center absorptions at 451 nm, matching the spectral
feature on Europa, rather than at the 460 nm wavelength
seen in previous experiments. Absorption at 460 nm is, however,
seen when samples are placed on the cold finger without pressing
into the foil and when thermally-coupled samples are heated
to 240 K. We conclude that previous experiments had poor thermal coupling between the cold finger and the angular salt grains  
resting on a cold finger, resulting in a raised sample temperature of approximately 240K. At the appropriate temperature, 
the band position of laboratory-irradiated NaCl at 451 nm falls within the uncertainty of the band position on Europa.

Previous laboratory data also predicted the presence of M-center
absorption at 720 nm on Europa for typical irradiation
doses
expected, but the HST data put strong upper limits on the
existence of such an absorption. Here we found that photobleaching
-- the destruction of F-centers by photons -- can nearly
precisely balance the irradiation-induced creation of F-centers,
leading to a low equilibrium value for F-center density
and a weak absorption strength,
similar to that seen on Europa. This low equilibrium
density is well below the density required for pairs of F-centers 
to begin to coalesce to form M-centers, explaining the
lack of this additional absorption on Europa.

Our model for F-center creation and destruction predicts that
the absorption will grow overnight in the absence of photobleaching
and partially decay during the day under solar illumination. We show that morning and
afternoon spectra of Tara Regio, when examined separately,
show this effect at a similar level as predicted.

These new laboratory experiments show that the visible
spectrum of Tara Regio is consistent with
the presence of irradiated NaCl, including
the wavelength of the F-center feature, the lack of an M-center
absorption at 720 nm, and the change in the strength of
the F-center feature over the course of the day.
The existence of NaCl in a chaos region on the leading
hemisphere of Europa strongly suggests that sodium and chlorine
are dominant components of the subsurface of Europa.

\acknowledgements
This research was supported by a grant number 668346 from the Simons Foundation.
The observations were made with the NASA/ESA
Hubble Space Telescope, obtained at the Space Telescope Science Institute, which is operated by the Association of Universities for Research in Astronomy, Inc.,
under NASA contract NAS5-26555. These observations
are associated with program $\#$14650. This work is
also associated with archival program $\#$15789.  This data was obtained from the Mikulski Archive for Space Telescopes (MAST). This specific observations analyzed can be accessed via\dataset[10.17909/t9-awev-6407]{https://doi.org/10.17909/t9-awev-6407}. Support for program $\#$15789 was provided by
NASA through grants from the Space Telescope Science
Institute, which is operated by the Association of Universities for Research in Astronomy, Inc., under NASA
contract NAS5-26555. S.K.T. is supported by the Heising-Simons Foundation through a \textit{51 Pegasi b} postdoctoral fellowship.

\end{document}